\documentclass[manuscript,screen]{acmart}
\AtBeginDocument{%
  \providecommand\BibTeX{{%
    \normalfont B\kern-0.5em{\scshape i\kern-0.25em b}\kern-0.8em\TeX}}}

\usepackage{caption}
\usepackage{subcaption}
\usepackage{comment}
\usepackage[T1]{fontenc}
\usepackage{arydshln}

\usepackage{enumitem}
\setlist[itemize]{leftmargin=*}

\begin{document}

\title{Do LLMs Understand User Preferences? Evaluating LLMs On User Rating Prediction}

\author{Wang-Cheng Kang*}{\email{wckang@google.com}\affiliation{
  \institution{Google Research, Brain Team}
    \country{USA}
}}
\author{Jianmo Ni*}{\email{jianmon@google.com}\affiliation{
  \institution{Google Research, Brain Team}
    \country{USA}
}}
\author{Nikhil	Mehta}{\email{nikhilmehta@google.com}\affiliation{
  \institution{Google Research, Brain Team}
    \country{USA}
}}
\author{Maheswaran Sathiamoorthy}{\email{nlogn@google.com}\affiliation{
  \institution{Google Research, Brain Team}
    \country{USA}
}}
\author{Lichan Hong}{\email{lichan@google.com}\affiliation{
  \institution{Google Research, Brain Team}
    \country{USA}
}}
\author{Ed Chi}{\email{edchi@google.com}\affiliation{
  \institution{Google Research, Brain Team}
    \country{USA}
}}
\author{Derek Zhiyuan Cheng}{\email{zcheng@google.com}\affiliation{
  \institution{Google Research, Brain Team}
    \country{USA}
}}

\thanks{*The two authors contributed equally to this work.}

\renewcommand{\shortauthors}{W.-C. Kang, et al.}
\begin{abstract}
  Large Language Models (LLMs) have demonstrated exceptional capabilities in generalizing to new tasks in a zero-shot or few-shot manner. However, the extent to which LLMs can comprehend user preferences based on their previous behavior remains an emerging and still unclear research question. Traditionally, Collaborative Filtering (CF) has been the most effective method for these tasks, predominantly relying on the extensive volume of rating data. In contrast, LLMs typically demand considerably less data while maintaining an exhaustive world knowledge about each item, such as movies or products. In this paper, we conduct a thorough examination of both CF and LLMs within the classic task of user rating prediction, which involves predicting a user's rating for a candidate item based on their past ratings. We investigate various LLMs in different sizes, ranging from 250M to 540B parameters and evaluate their performance in zero-shot, few-shot, and fine-tuning scenarios. We conduct comprehensive analysis to compare between LLMs and strong CF methods, and find that zero-shot LLMs lag behind traditional recommender models that have the access to user interaction data, indicating the importance of user interaction data. However, through fine-tuning, LLMs achieve comparable or even better performance with only a small fraction of the training data, demonstrating their potential through data efficiency.
\end{abstract}

\maketitle

\section{Introduction}

Large language models (LLMs) have shown an uncanny ability to handle a wide variety of tasks such as text generation \cite{chowdhery2022palm, devlin2018bert, colin_t5, brown2020language}, translation \cite{wang2019learning, yang2020towards}, and summarization \cite{liu2019text}. The recent fine-tuning of LLMs on conversations and the use of techniques like instruction fine-tuning \cite{flan-t5} and reinforcement learning from human feedback (RLHF) \cite{rlhf} led to tremendous success to bring highly human-like chatbots (e.g. ChatGPT \cite{chatgpt}, and Bard \cite{bard}) to the average household.
There are three key factors that directly contribute to LLMs' versatility and effectiveness:

\begin{enumerate}

    \item Knowledge from internet-scale real world information: LLMs are trained on enormous datasets of text, providing access to a wealth of real-world information. This information is converted to knowledge that can be used to answer questions, creatives writing (e.g., poems, and articles), and translate between languages.
    \item Incredible generalization ability through effective few-shot learning: within certain context, LLMs are able to learn new tasks from an extremely small number of examples (a.k.a., few-shot learning). Strong few-shot learning capability gears up LLMs to be highly adaptable to new tasks.
    \item Strong reasoning capability: LLMs are able to reason through a chain-of-thought process \cite{wei2022chain, wei2022emergent}, significantly improving their performance across many tasks \cite{wang2023selfconsistency}.
\end{enumerate}

Recently, there has been some early exploratory work to make use of LLMs for Search \cite{bing_chatgpt}, Learning to Rank \cite{kdd2021_baidu, han2020learningtorank}, and Recommendation Systems \cite{geng2022recommendation, cui2022m6, liu2023chatgpt}. Specifically for recommendation systems, P5~\cite{geng2022recommendation} fine-tunes T5-small (60M) and T5-base(220M)~\cite{raffel2020exploring}, unifying both ranking, retrieval and other tasks like summary explanation into one model.
M6-Rec~\cite{cui2022m6} tackles the CTR prediction task by finetuning a LLM called M6~(300M)~\cite{lin2021m6}.
Liu et al. \cite{liu2023chatgpt} looked into whether conversational agents like ChatGPT could serve as an off-the-shelf recommender model with prompts as the interface and reported zero-shot performance on rating prediction against baselines like MF and MLP. However, there is a noticeable absence of a comprehensive study that meticulously evaluates LLMs of varying sizes and contrasts them against carefully optimized, strong baselines.

In this paper, we explore the use of off-the-shelf large language models (LLMs) for recommendation systems. We study a variety of LLMs of various sizes ranging from 250M to 540B parameters. We focus on the specific task of user rating prediction, and evaluate the performance of these LLMs under three different regimes: 1. zero-shot 2. few-shot, and 3. fine-tuning. We then carefully compare them with the state-of-the-art recommendation models on two widely adopted recommendation benchmark datasets.
\begin{figure}[t]
\centering
\includegraphics[width=0.9\textwidth]{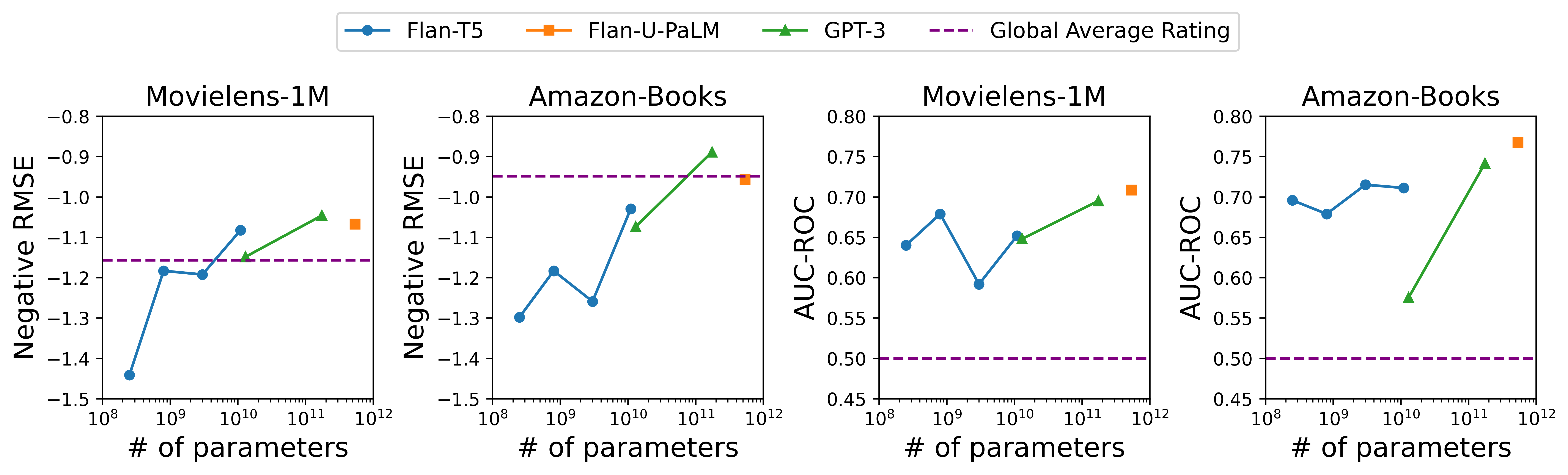}
\caption{Zero-shot performance for rating prediction of LLMs with different model sizes, including Flan-T5 (base to XXL), GPT3 (Curie, Davinci) and Flan-U-PaLM. We see a performance gain when increasing the model size. Among which, models greater than 100B (Flan-U-PaLM 540B and text-davinci-003 175B) outperform or on-par with the global average rating baseline on both RMSE and AUC-ROC.}
\vspace{-1em}
\label{fig:zero-shot}
\end{figure}
Our contributions are three-fold:
\begin{itemize}
\item We empirically study the zero-shot and few-shot performance of off-the-shelf LLMs with a wide spectrum of model sizes. We found that larger models (over 100B parameters) can provide reasonable recommendations under the cold-start scenario, achieving comparable performance to decent heuristic-based baselines.
\item We show that zero-shot LLMs still fall behind traditional recommender models that utilize human interaction data. Zero-shot LLMs only achieve comparable performance than two surprisingly trivial baselines that always predicts the average item or user rating. Furthermore, they significantly underperform traditional supervised recommendation models, indicating the importance of user interaction data.
\item Through numerous experiments that fine-tune LLMs on human interaction data, we demonstrate that fine-tuned LLMs can achieve comparable or even better performance than traditional models with only a small fraction of the training data, showing its promise in data efficiency.
\end{itemize}

\section{Related Work}

\subsection{Use of Natural Language in Recommender System}
One of the earliest works that explored formulating the recommendation problem as a natural language task is ~\cite{zhang2021language}. They used BERT~\cite{devlin2018bert} and GPT-2~\cite{radford2019language} on the Movielens dataset~\cite{harper2015movielens} to show that such language models perform surprisingly well, though not as good as well tuned baselines like GRU4Rec~\cite{hidasi2015session}.

P5~\cite{geng2022recommendation} fine-tunes a popular open-sourced  T5~\cite{raffel2020exploring} model, unifying both ranking, retrieval and other tasks like summary explanation into one model. 
M6-Rec~\cite{cui2022m6} is another related work, but they tackle the CTR prediction task by finetuning a LLM called M6~\cite{lin2021m6}.

Two recent works explore the use of LLMs for zero-shot prediction. ChatRec~\cite{gao2023chat} handles zero-shot prediction as well as being interactive and providing explanations. \cite{wang2023zero} takes a three-stage prompting approach to generate next item recommendation in the Movielens dataset and achieves competitive metrics, although not being able to beat strong sequential recommender baselines such as SASRec~\cite{kang2018self}.

\subsection{Large Language Models}
Once people realized that scaling up sizes of data and model helps language models, there has been a series of large language models proposed and built: e.g. PaLM~\cite{chowdhery2022palm}, GPT-3~\cite{brown2020language} and recent ones such as OPT~\cite{zhang2022opt} and LLaMA~\cite{touvron2023llama}. One of the unique abilities of LLMs has been in their ability to reason about things, which is further improved by techniques such as chain-of-thought prompting~\cite{wei2022chain}, self-consistency~\cite{wang2022self} and self-reflection~\cite{shinn2023reflexion}. 

Another major strong capability of LLMs is instruction following that models can generalize to unseen tasks by following the given natural language instructions. Researchers have found that techniques like instruction fine-tuning \cite{flan-t5} and RLHF \cite{rlhf} can significantly improve LLMs' capability to perform tasks given natural language descriptions that align with human's preferences. As one of the tasks that can be described in natural language, `recommendation' has become a promising new capability for LLMs. In this work, we focus on the models that have been fine-tuned to improve their instruction following capability such as ChatGPT \cite{chatgpt}, GPT-3 (text-davinci-003 \cite{instructgpt}), Flan-U-PaLM and Flan-T5 \cite{flan-t5}.

\section{Method}
\subsection{Problem formulation}

We study the task of user rating prediction, formulated as: Given a user $u \in \mathcal{U}$, a sequence of user $u$'s historical interactions $E^u=\{e^u_1, e^u_2, ..., e^u_{n}\}$ and an item $i \in \mathcal{I}$, predict the rating that the user $u$ will give to the item $i$, where the user historical interaction sequence $E^u$ is ordered by time ($e^u_n$ is the most recent item that the user consumed), and each interaction $e^u_k$ is represented by information about the item (e.g., ID, title, metadata, etc.) that the user has consumed as well as the rating the user gave to the item.

\subsection{Zero-shot and Few-shot LLMs for Rating Prediction}
\begin{figure}[h]
\vspace{-0.2cm}
 \centering
\centering
\includegraphics[width=\textwidth]{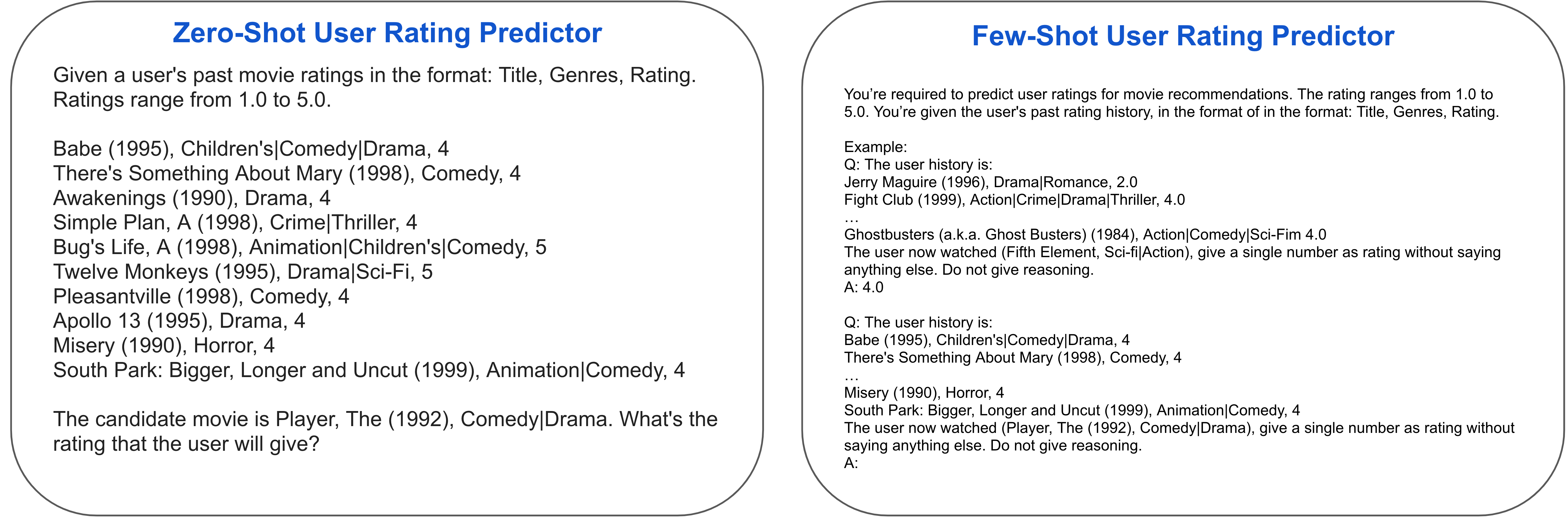}
\caption{Zero and few shot LLM prompt for rating prediction.}
\label{fig:prompt}
\end{figure}
We demonstrate the zero-shot and few-shot prompts used for the rating prediction task on the MovieLens dataset in figure~\ref{fig:prompt}. As shown in the figure, the input prompts depict several important features represented as text, including user's past rating history and candidate item features (title and genre). Finally, to elicit a numeric rating from the model with the rating scale, the input prompt specifies a numerical rating scale. The model response is parsed to extract the rating output from the model. However, we discovered that LLMs can be highly sensitive to the input prompts and do not always follow the provided instruction. For instance, we found that certain LLMs may offer additional reasoning or not provide a numerical rating at all. To resolve this, we performed additional prompt engineering by adding additional instructions such as "Give a single number as rating without explanation" and "Do not give reasoning" to the input prompt.

\subsection{Fine-tuning LLMs for Rating Prediction}

In traditional recommender system research, it has been widely shown that training models with human interaction data is effective and critical to improve recommender's capability of understanding user preference. 

\begin{figure}
 \centering
 
 \begin{subfigure}{0.2\textwidth}
 \small
     \centering
     \includegraphics[width=0.8\textwidth]{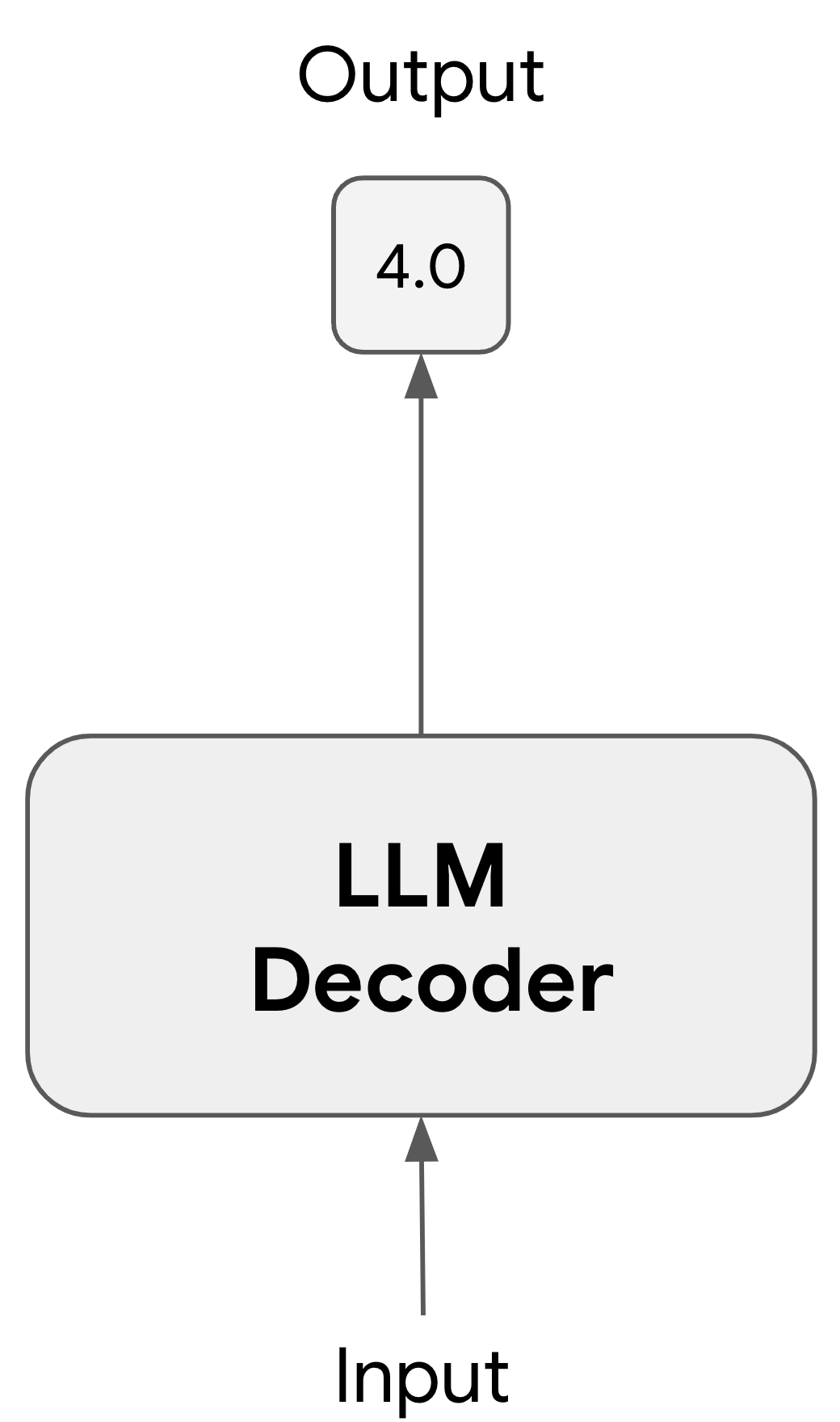}
     \caption{Decoder-Only model.}
     \label{fig:dec-only}
 \end{subfigure}
\hspace{2em}
 \begin{subfigure}{0.7\textwidth}
 \small
     \centering
     \includegraphics[width=0.8\textwidth]{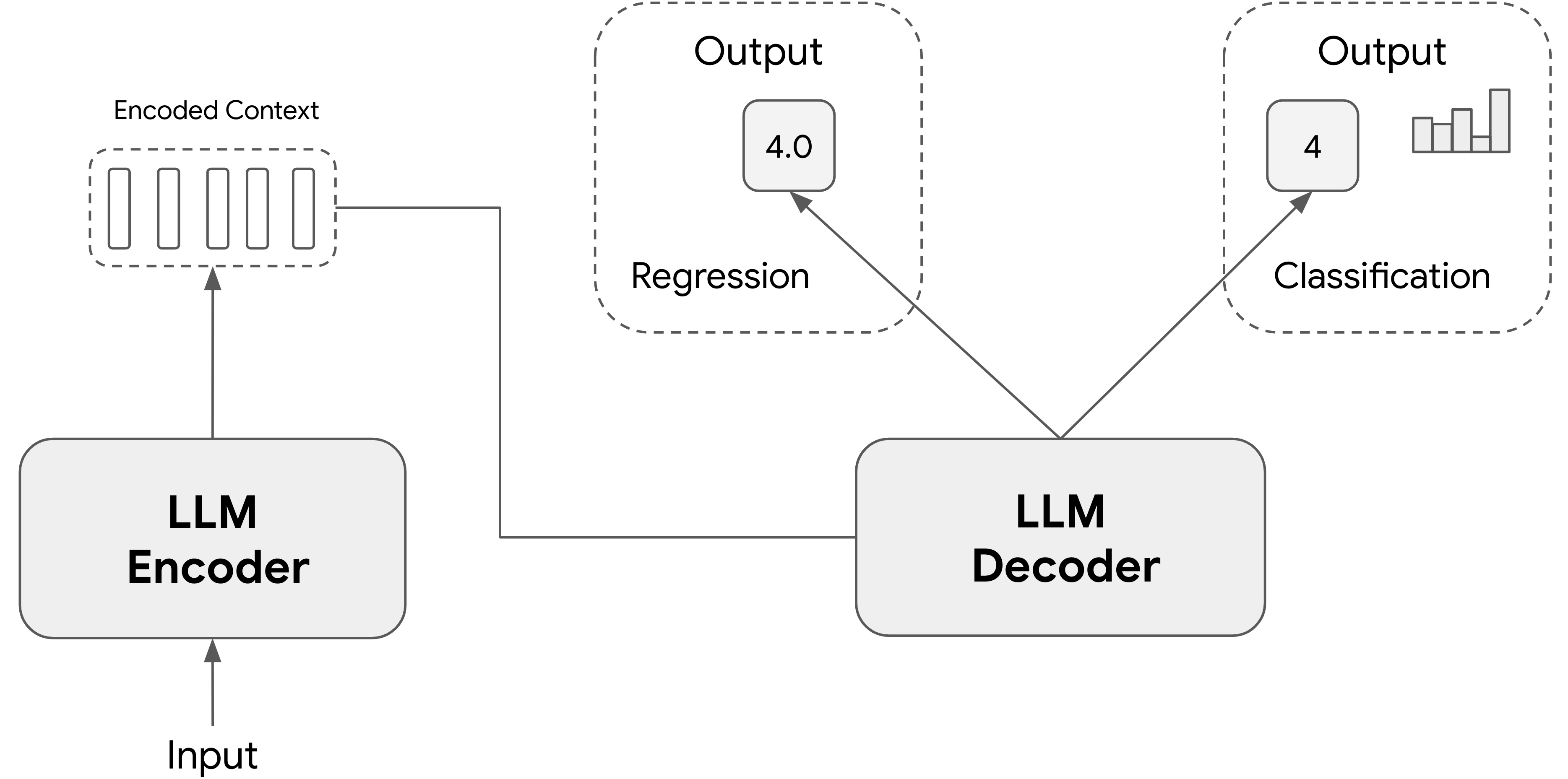}
     \caption{Encoder-Decoder model.}
     \label{fig:encoder-decoder}
 \end{subfigure}
    \caption{Two types of LLMs for the rating prediction task.}
    \label{fig:llm-model}
\end{figure}

Here, we explore training the LLMs with human interaction and study how it could improve the model performance. We focus on fine-tuning a family of LLMs, namely Flan-T5, since they are publicly available and have competitive performance on a wide range of benchmarks. The rating prediction task could be formulated into one of two tasks: (1) multi-class classification; or (2) regression, as shown in Figure \ref{fig:encoder-decoder}.

\paragraph{Multi-class Classification} LLMs (either Decoder-only or Encoder-Decoder architecture) are essentially pre-trained with a $K$-way classification task that predicts the token from a fixed vocabulary with size $K$. As shown in Figure \ref{fig:encoder-decoder}, there is a projection layer that projects the outputs from the last layer to the vocabulary size then the pre-training optimizes the cross-entropy loss for the token classification. The output logits is computed as $\mathrm{logits}_{\mathrm{dec}} = W_{\mathrm{proj}} h_{\mathrm{dec}}$,
where $W_{\mathrm{proj}}$ is the projection matrix of size $(d, |V|)$, $h_{\mathrm{dec}}$ is the output from the decoder's last transformer layer, $d$ is the hidden dimension size of the decoder and $|V|$ is the vocabulary.  

Following \cite{colin_t5,flan-t5}, we formulate the rating regression task as a 5-way classification task, where we take the rating 1 to 5 as 5 classes. 
During training, we use the cross-entropy loss as other classification tasks, as shown below:
\begin{equation}
    L_{\mathrm{cross\_entropy}} = - \sum_{i=1}^{N} r^i \mathrm{log}(\mathrm{logits}_{\mathrm{dec}}^i),
\label{eq:cross-entropy}
\end{equation}
where $r^i$ is the ground-truth rating for the $i$-th item and $N$ is the number of total training examples.

During inference, we compute the log-likelihood for the model output each class and choose the class with the largest probability as the final prediction.

\paragraph{Regression}. To enable LLMs for regression tasks, we set the shape of the projection matrix $W_{\mathrm{proj}}$ to be $(d, 1)$, so that it will only output a 1-digit logits. As shown in Equation \ref{eq:regression}, during training we apply a mean-squared-error (MSE) loss based on the output logits and the ground-truth rating.

\begin{equation}
    L_{\mathrm{regression}} = \frac{1}{|N|} \sum_{i=1}^{N} (\mathrm{logits}_{\mathrm{dec}}^i - r^i)^2.
\label{eq:regression}
\end{equation}

\section{Experiments}
\newcommand{\B}{\textbf}
\newcommand{\UL}[1]{\underline{#1}}

We conduct extensive experiments to answer the following research questions:\newline
\textbf{RQ1}: Do off-the-shelf LLMs perform well for zero-shot and few-shot recommendations? \newline
\textbf{RQ2}: How do LLMs compare with traditional recommenders in a fair setting \newline
\textbf{RQ3}: How much does model size matter for LLMs when used for recommenders? \newline
\textbf{RQ4}: Do LLMs converge faster than traditional recommender models?

\subsection{Datasets and Evaluation Setup}

\subsubsection{Datasets} To evaluate the user rating prediction task, we use two widely adopted benchmark datasets for evaluating model performance on recommendations. Both datasets consist of user review ratings that range from 1 to 5.
\begin{itemize}
    \item MovieLens \citep{Harper2016TheMD}: We use the version MovieLens-1M that includes 1 million user ratings for movies.
    \item Amazon-Books \citep{Ni2019JustifyingRU}: We use the ``Books'' category of the Amazon Review Dataset with users' ratings on items. We use the 5-core version that filters out users and items with less than 5 interactions.
\end{itemize}

\subsubsection{Training / Test Split} To create the training and test sets, we follow the single-time-point split \cite{Sun2022TakeAF}. We first filter out the ratings associated with items that don't have metadata, then sort all user ratings in chronological order. Finally, we take the first 90\% ratings as the training set and the remaining as the test set. Each training example is a tuple of $<user\_id, item\_id, item\_metadata, rating>$, where the label is a 5 Likert-scale rating. The input features are $user\_id$, $item\_id$, and a list of item\_metadata features. The statistics of the datasets are shown in the Table \ref{tab:dataset_stats}.
Due to the high computation cost of zero-shot and few-shot experiments based on LLMs, we randomly sample from the test set of each dataset into 2,000 tuples as a smaller test set. For all our experiments, we report results on the sampled test set. And we truncate the user sequence to the most recent 10 interactions during training and evaluation.

\begin{table}
\caption{Statistics of the dataset}
\label{tab:dataset_stats}
\begin{tabular}{l|r|r|r|r|c}
\toprule
Datasets & \#Users & \#Items & \# of Training Examples & \# of Test Examples & Features \\
\midrule
Movielens-1M & 6,040 & 3,689 & 882,002 & 2,000 (75,880) & Title, Genre \\
Amazon-Books & 1,850,187 & 483,579 & 17,566,711 & 2,000 (2,324,503) & Title, Brand \\
\bottomrule
\end{tabular}
\end{table}

\subsubsection{Evaluation Metrcis} We use the widely adopted metrics RMSE (Root Mean Squared Error) and MAE (Mean Average Error) to measure model performance on rating prediction. Moreover, we use ROC-AUC to evaluate the model's performance on ranking, where ratings greater than or equal to 4 are considered as positives and the rest as negatives. In this case, AUC measures whether the model ranks the positives higher than negatives.

\subsection{Baselines and LLMs}

\subsubsection{Baselines}
\begin{itemize}

\item \textbf{Traditional Recommeder}: We consider several traditional recommendation models as strong baselines, including 1. Matrix Factorization~(\textbf{MF}) \cite{rendle2012bpr}, and 2. Multi-layer Perceptrons~(\textbf{MLP})~\cite{DBLP:conf/www/HeLZNHC17}. For MF and MLP, only user ID and item ID are used as input features.

\item \textbf{Attribute and Rating-aware Sequential Rating Predictor}: In our experiments, we supply the LLM with historical item metadata, such as titles and categories, along with historical ratings. However, to the best of our knowledge, there is no existing method designed for this setting\footnote{The most related works are SASRec~\cite{kang2018self} and CARCA~\cite{rashed2022carca}, however they are designed for next item prediction instead of rating prediction, and thus not directly applicable to our case.}. To ensure a fair comparison, we construct a \mbox{\textbf{Transformer-MLP}}
model to efficiently process the same input information provided to the LLM.

There are three key design choices: (i) feature processing: We treat all features as sparse features, and learn their embeddings end-to-end. For example, we use one-hot encoding for genres, and create an embedding table, where the i-th row is genre i's embedding. Similarly, we obtain bag-of-words encodings via applying a tokenizer\footnote{\url{https://www.tensorflow.org/text/api_docs/python/text/WhitespaceTokenizer}} on titles, and then look up the corresponding embedding. (ii) user modeling: for each user behavior, we use \texttt{Add} or \texttt{Concat} to aggregate all embeddings (e.g. item ID, title, genres/category, rating) into one, and then adopt bi-directional self-attention~\cite{vaswani2017attention} layers with learned position embeddings to model users' past behaviors. Similar to SASRec, we use the most recent behavior's output embedding as the user summary; (iii) Fuse user and candidate for rating prediction: we apply a MLP on top of the user embedding along with other candidate item features to generate the final rating prediction, and optimize for minimizing MSE.
    
To properly tune the baseline models, we defined a hyper-parameter search space (e.g. for embedding dimension, learning rate, network size, \texttt{Add} or \texttt{Concat} aggregation, etc.), and perform more than 100 search trials using Vizier \cite{google_vizier}, a black-box hyper-parameter optimization tool.

\item \textbf{Heuristics}: We also include three heuristic-based baselines: (1) global average rating: (2) candidate item average rating, and (3) user past average rating, meaning the model's prediction is depending on (1) the average rating among all user-item ratings, (2) the average rating from the candidate item or (3) the user's average rating in the past.
\end{itemize}

\subsubsection{LLMs for Zero-shot and Few-shot Learning}: We used the LLMs listed below for zero-shot and few-shot learning. We use a temperature of 0.1 for all LLMs, as the LLM's output in our case is simply a rating prediction. We use GPT-3 models from OpenAI~\cite{OpenAI_models}: (i) \textbf{text-davinci-003 (175B)}: The most capable GPT-3 model with Reinforcement Learning from Human Feedback (RLHF)~\cite{DBLP:journals/corr/abs-2009-01325}; (ii) \textbf{ChatGPT}: the default model is gpt-3.5-turbo, fine-tuned on both human-written demonstrations and RLHF, and further optimized for conversation. \textbf{Flan-U-PaLM (540B)} is the largest and strongest model in~\cite{flan-t5}, it applies both FLAN instruction tuning~\cite{DBLP:conf/iclr/WeiBZGYLDDL22} and UL2 training objective~\cite{flan-t5} on PaLM~\cite{chowdhery2022palm}.

\begin{table}
\small
\caption{User rating prediction results. The best performing method is boldfaced in each column, and underlined in each group.}
\label{tab:main}
\begin{tabular}{lcccccc}
\toprule
Model & \multicolumn{3}{c}{MovieLens} & \multicolumn{3}{c}{Amazon-Books}\\
                    &RMSE$\downarrow$  & MAE$\downarrow$  & AUC$\uparrow$&RMSE$\downarrow$  & MAE$\downarrow$  & AUC$\uparrow$ \\
\midrule
\emph{Zero-Shot LLMs}\\ 
Flan-U-PALM         &1.0677     &\UL{0.7740}&\UL{0.7084}&0.9565     &0.5569     &\UL{0.7676}\\
ChatGPT       &\UL{1.0081}&0.8193     &0.6794     &1.0081     &0.8093     &0.6778\\
text-davinci-003    &1.0460     &0.7850     &0.6951     &\UL{0.8890}&\UL{0.5442}     &0.7416\\
\hdashline
\emph{Few-Shot LLMs}\\
Flan-U-PALM         &\UL{1.0721}     &\UL{0.7605}     &\UL{0.7094}     &1.0712     &\UL{0.5855}     &0.7439\\
ChatGPT       &1.0862     &0.8203     &0.6930     &\UL{1.0618}     &0.7760     &0.7470\\
text-davinci-003    &1.0867     &0.8119     &0.6963     &1.0716 &0.7753     &\UL{0.7739}\\
\hdashline
\emph{Simple Dataset Statistics}\\ 
Global Avg. Rating       &1.1564     &0.9758       &0.5        &0.9482     &0.7609     &0.5\\
Candidate Item Avg. Ratings   &\UL{0.9749}&\UL{0.7778}&\UL{0.7395}&0.9342     &0.7078     &0.6041\\
User Past Avg. Ratings   &1.0196     &0.7959     &0.7266     &\UL{0.8527}&\UL{0.5502}&\UL{0.8047}\\
\hdashline
\emph{Supervised Recommendation Methods}\\ 
MF                      &0.9552     &0.7436     &0.7734     &1.7960     &1.1070    &0.7638\\
MLP                     &0.9689     &0.7452     &0.7393     &0.8607     &0.6384     &0.6932\\
Transformer+MLP         &\UL{\B{0.8848}} &\UL{0.7036}     &\UL{0.7979}     &\UL{\B{0.8143}} &\UL{0.5541}     &\UL{0.8042} \\
\hdashline
\emph{Fine-tuned LLMs}\\ 
Flan-T5-Base (classification) & 1.0110 & 0.6805 & 0.7590 & 0.9856 & 0.4685 & 0.6292 \\
Flan-T5-Base (regression)        &0.9187     &0.7092     &0.7949     &0.8413     &0.5317     &0.8182\\
Flan-T5-XXL (regression)        &\UL{0.8979}     &\UL{\B{0.6986}} &\UL{\B{0.8042}} &\UL{0.8301}     &\B{\UL{0.5122}} &\UL{\B{0.8312}}\\
\bottomrule
\end{tabular}
\vspace{-1em}
\end{table}

\begin{figure}[t]
 \centering
 \begin{subfigure}{0.49\textwidth}
 \small
     \centering
     \includegraphics[width=\textwidth]{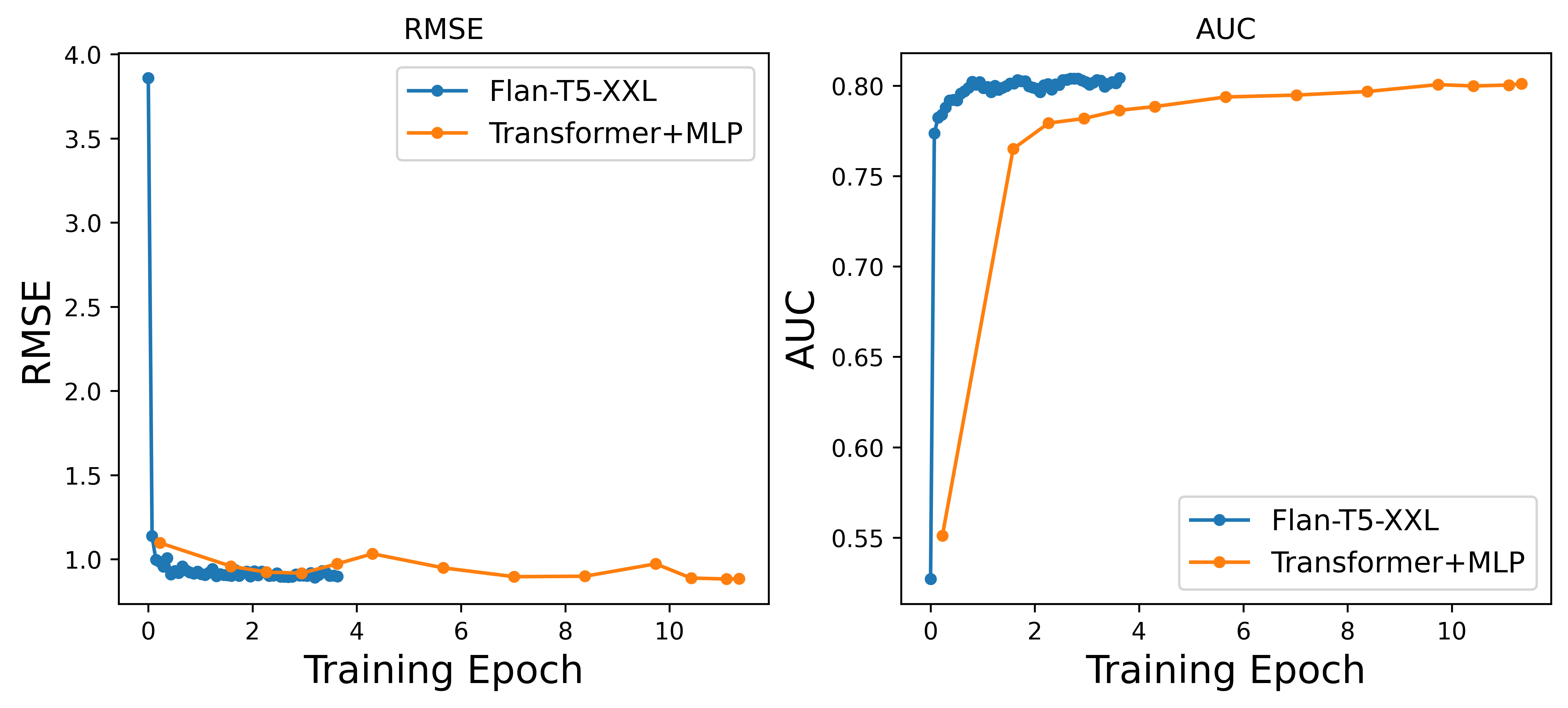}
     \caption{MovieLens}
 \end{subfigure}
 \hfill
 \begin{subfigure}{0.49\textwidth}
 \small
     \centering
     \includegraphics[width=\textwidth]{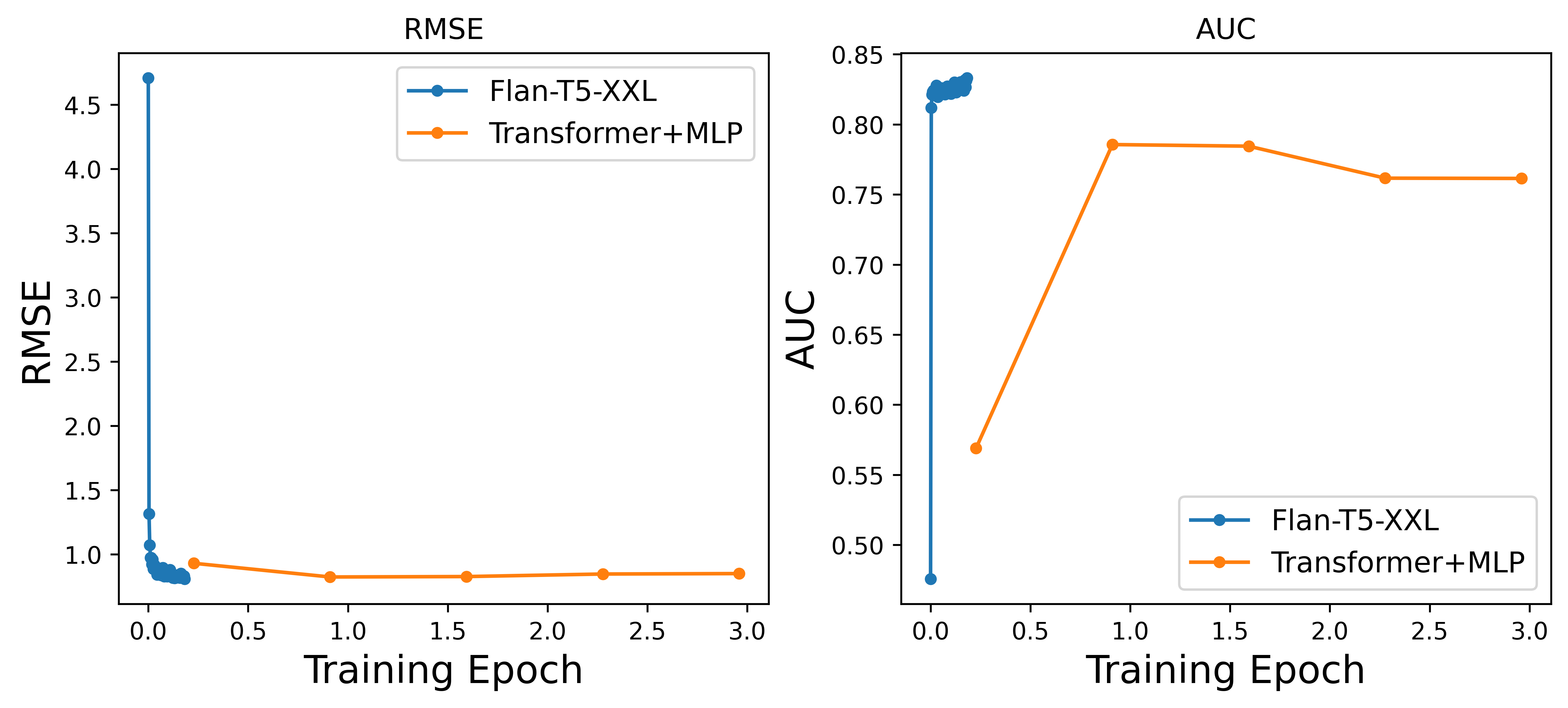}
     \caption{Amazon Books}
 \end{subfigure}
    \caption{Data efficiency: convergence curve.}
    \label{fig:exp:data_efficiency}
\end{figure}

\subsubsection{LLMs for Fine-tuning} For fine-tuning methods, we use \textbf{Flan-T5-Base} (250M) and \textbf{Flan-T5-XXL} (11B) models in the experiments. We set the learning rate to 5e-5, batch size to 64, drop out rate to 0.1 and train 50k steps on all datasets.

\subsection{Zero-Shot and Few-shot LLMs (RQ1)}

As shown in Table~\ref{tab:main}, we conduct experiments on several off-the-shelf LLMs in the zero-shot setting. We observed that LLMs seem to understand the task from the prompt description, and predict reasonable ratings. LLMs outperform global average rating in most cases, and perform comparable with item or user average ratings. For example, text-davinci-003 performs slightly worse than candidate item average ratings on Movielens but outperforms on Amazon-Books. %
For few-shot experiments, we provide 3 examples in the prompt (3-shot).
Compared against zero-shot, we found that the AUC for few-shot LLMs are improved, while there is no clear pattern in RMSE and MAE.

Furthermore, we found that both zero-shot and few-shot LLMs under-perform traditional recommendation models trained with interaction data. As shown in Table \ref{tab:main}, GPT-3 and Flan-U-PaLM models achieve significantly lower performance compared to supervised models. The inferior performance could be due to the lack of user-item interaction data in LLMs' pre-training, and hence they don't have knowledge about human preference for different recommendation tasks. 
Moreover, recommendation tasks are highly dataset-dependent: (e.g.) the same movie can have different average ratings on different platform. Hence, without knowing the dataset-specific statistics, it's impossible for LLM to provide a universal prediction that is suitable for every datasets.

\subsection{LLMs vs. Traditional Recommender Models (RQ2)}

Fine-tuning LLMs is an effective way to feed dataset statistics into LLMs, and we found the performance of fine-tune LLMs are much better than zero/few-shot LLMs. Also, when fine-tuning the Flan-T5-base model with the classification loss, the performance is much worse than fine-tuning with the regression loss on all three metrics. This indicates the importance of choosing the right optimizing objective for fine-tuning LLMs.

Comparing against the strongest baseline Transformer-MLP, we found fine-tuned Flan-T5-XXL has better MAE and AUC, implying fine-tune LLMs may be more suitable for ranking tasks.

\subsection{Effect of Model Size (RQ3)}

For all the LLMs we studied of different model sizes vary from 250M and 500B parameters, we were able to use zero-shot or few-shot prompts to let them output a rating prediction between 1 to 5. This shows the effectiveness of instruction tuning that enables these LLMs (Flan-T5, Flan-U-PaLM, GPT-3) to follow the prompt. We further found that only LLMs with size greater than 100B perform reasonably well on rating prediction in the zero-shot setting, as shown in Figure \ref{fig:zero-shot}. For fine-tuning experiments, we also found that Flan-T5-XXL outperforms Flan-T5-Base on both datasets, as shown in the last two rows of Table \ref{tab:main}.

\subsection{Data Efficiency of LLMs (RQ4)}

As LLMs have learned vast amounts of world knowledge during pre-training, while traditional recommender models are trained from scratch, we compare their convergence curves in Figure \ref{fig:exp:data_efficiency} to examine whether LLMs have better data efficiency. We can see that for RMSE, both methods could converge to reasonable performance with a small fraction of data. This is probably because that even average rating of all items has a relatively low RMSE, and thus as long as a model learns to predict a rating near the average rating, it could achieve reasonable performance. For AUC the trend is more clear, as simply predicting average rating results in an AUC of 0.5. We found that a small fraction of data is required for LLM to achieve good performance, while Transformer+MLP needs much more training data (at least 1 epoch) for convergence.

\section{Conclusion}
In this paper, we evaluate the effectiveness of large language models as a recommendation system for user rating prediction in three settings: 1. zero-shot; 2. few-shot; and 3. fine-tuning. 
Compared to traditional recommender methods, our results revealed that LLMs in zero-shot and few-shot LLMs fall behind fully supervised methods, implying the importance of incorporating the target dataset distribution into LLMs. On the other hand, fine-tuned LLMs can largely close the gap with carefully designed baselines in key metrics. LLM-based recommenders have several benefits: (i) better data efficiency; (ii) simplicity for feature processing and modeling: we only need to convert information into a prompt without manually designing feature processing strategies, embedding methods, and network architectures to handle various kind of information; (iii) potential for unlock conversational recommendation capabilities. Our work sheds light on the current status of LLM-based recommender systems, and in the future we will further look into improving the performance via methods like prompt tuning, and explore novel recommendation applications enabled by LLMs.

\bibliographystyle{ACM-Reference-Format}
\bibliography{references}

\end{document}